\documentclass[conference]{IEEEtran}
\IEEEoverridecommandlockouts
\usepackage{cite}
\usepackage{amsmath,amssymb,amsfonts}
\usepackage{algorithmic}
\usepackage{graphicx}
\usepackage{textcomp}
\usepackage{xcolor}
\usepackage{array}
\usepackage{multirow}
\usepackage{comment}

\def\BibTeX{{\rm B\kern-.05em{\sc i\kern-.025em b}\kern-.08em
    T\kern-.1667em\lower.7ex\hbox{E}\kern-.125emX}}
    
\usepackage{fancyhdr}
\begin{document}

\title{Privileged Information for Modeling Affect\\ In The Wild\\
{}
\thanks{This work has been supported by the European Union’s Horizon 2020 research and innovation programme from the TAMED project (Grant Agreement No. 101003397).}
}

\author{\IEEEauthorblockN{Konstantinos Makantasis}
\IEEEauthorblockA{\textit{Institute of Digital Games} \\
\textit{University of Malta}\\
Msida, Malta \\
konstantinos.makantasis@um.edu.mt}
\and
\IEEEauthorblockN{David Melhart}
\IEEEauthorblockA{\textit{Institute of Digital Games} \\
\textit{University of Malta}\\
Msida, Malta \\
david.melhart@um.edu.mt}
\and
\IEEEauthorblockN{Antonios Liapis}
\IEEEauthorblockA{\textit{Institute of Digital Games} \\
\textit{University of Malta}\\
Msida, Malta \\
antonios.liapis@um.edu.mt}
\and
\IEEEauthorblockN{Georgios N. Yannakakis}
\IEEEauthorblockA{\textit{Institute of Digital Games} \\
\textit{University of Malta}\\
Msida, Malta \\
georgios.yannakakis@um.edu.mt}
}

\maketitle
\thispagestyle{fancy}

\begin{abstract}
A key challenge of affective computing research is discovering ways to reliably transfer affect models that are built in the laboratory to real world settings, namely \emph{in the wild}. The existing gap between \emph{in vitro} and \emph{in vivo} affect applications is mainly caused by limitations related to affect sensing including intrusiveness, hardware malfunctions, availability of sensors, but also privacy and security. As a response to these limitations in this paper we are inspired by recent advances in machine learning and introduce the concept of \emph{privileged information} for operating affect models in the wild. The presence of privileged information enables affect models to be trained across multiple modalities available in a lab setting and ignore modalities that are not available in the wild with no significant drop in their modeling performance. The proposed privileged information framework is tested in a game arousal corpus that contains physiological signals in the form of heart rate and electrodermal activity, game telemetry, and pixels of footage from two dissimilar games that are annotated with arousal traces. By training our arousal models using all modalities (in vitro) and using solely pixels for testing the models (in vivo), we reach levels of accuracy obtained from models that fuse all modalities both for training and testing. The findings of this paper make a decisive step towards realizing affect interaction in the wild.
\end{abstract}


\section{Introduction}
Enabling forms of affective interaction in real-world applications and settings, i.e. \emph{in the wild}, is a core vision for affective computing (AC) \cite{kollias2019deep,toisoul2021estimation}. AC in the wild, however, remains a critical challenge as AC systems required to operate outside a controlled environment are limited by the accuracy of their affect models. Most importantly, the performance of such systems is determined largely by the quality of affect sensing in the wild.

Sensing affect in real-world settings is limited by a number of factors. First, it is often the case that sensing equipment (e.g. cameras, physiological sensors, microphones) either introduce data bias due to environmental conditions or experimental noise due to hardware failure. Naturally, if the prediction of affect is dependent on such modalities, the affect models will under-perform. Second, as AC occurs \emph{in vivo}, sensors and multimodal sensing information are often \emph{not available}, e.g., in users' homes or in their cars \cite{eyben2010emotion,braun2019improving}, and in public spaces including museums \cite{kim2015measuring}, hospitals or rehabilitation centers \cite{tripathi2021advancing}, which define some of the most popular application domains of AC. Finally, information about users in the wild often comes with a cost in terms of \emph{intrusiveness} (e.g. requiring users to use sensors) and \emph{privacy} (e.g. access to a smartphone's webcam and microphone). 

In this paper we are inspired by recent trends in machine learning and introduce the notion of \emph{privileged information} for modeling affect in an attempt to overcome the current limitations of affect sensing in the wild. In particular, the \emph{Learning Using Privileged Information} (LUPI) paradigm \cite{vapnik2009new, vapnik2015learning} is best suited for tasks featuring dissimilar amounts of information available during the model training and the model testing phases. Our hypothesis is that the LUPI paradigm can be beneficial for the performance of affect modeling; in particular sensing and operating these models in the wild. We test our hypothesis in the domain of video games by constructing models of player annotated arousal across two dissimilar games. Similarly to \cite{makantasis2019pixels,makantasis2021pixels}, our deep network models rely on gameplay pixels for both training and testing phases. We offer our models \emph{privileged information} in the form of game telemetry data and physiological signals---electrodermal activity and heart rate---during the training phase; these data types are normally available through extensive user tests in a quality assurance department or lab. We test our arousal models, however, without this privileged information, in the wild, as would be the case in a player's home setting. Our results suggest that the arousal models trained through privileged information perform equally well as the models trained with all modalities (pixels, telemetry and physiology) available to the model during both training and testing. Our findings suggest that privileged information is a critical milestone for realising AC in the wild as models trained on information that is available in a lab setting (\emph{in vitro}) can perform equally well when that information is not available or is distorted in the wild (\emph{in vivo}). Through privileged information the models of affect gain on unobtrusiveness, accessibility and practicality. 

This paper is novel in a number of ways. First, to the best of our knowledge, this study applies for the first time the LUPI paradigm in affective computing for building affect models that use different modalities of information during their training and testing phases. Second, following the LUPI paradigm, we present a rigorous methodology for deriving models of affect that perform equally well \textit{in vitro} and \textit{in vivo}, making a step towards accurate affect modeling in the wild. Finally, we validate our approach across two different games with regards to privileged (telemetry and physiology) and non-privileged information, as well as the arousal patterns they elicit. The validation results suggest that exploiting privileged information can boost the performance of affect models that operate in the wild.

\section{Related Work}

This section covers the related areas of pixel-based affect modeling and affect modeling in the wild.

\subsection{Pixel-based Affect Modeling}

Due to the richness of information encoded in videos and pictures, eliciting and modeling emotion via visual cues has been at the core of interest in affective computing \cite{picard2000affective}.

Before the deep learning era, the dominant approach for representing visual content was based on ad-hoc handcrafted features. Popular visual descriptors, such as SIFT \cite{lowe2004distinctive}, HOG \cite{dalal2005histograms} and LBP \cite{ahonen2004face} have been widely used to represent human facial patterns via a set of features used as inputs to machine learning models for emotion recognition \cite{ko2018brief,zheng2010emotion}. 
All those approaches are characterized by low computing power and memory requirements and, thus, are still being studied for use in real-time embedded systems \cite{suk2014real}. 

In the last decade, there has been a breakthrough in deep convolution neural networks (CNNs) applied to the field of computer vision. CNNs enable the end-to-end training directly from input images unifying, this way, the feature extraction and classification tasks. CNNs were first applied in \cite{baveye2015deep} to predict dimensional affective scores from videos. 
The need for effectively training deep learning models triggered the development of medium- and large-scale affect corpora \cite{zhang2012finding,zafeiriou2017aff}. Breuer and Kimmer in \cite{breuer2017deep} train CNNs and demonstrate their capacity to jointly learn various facial expression recognition tasks. In \cite{jung2015joint}, the authors use CNNs to learn geometric and temporal features describing facial action points to boost facial-based affect models' performance. Ng \textit{et al.} \cite{ng2015deep} propose transfer of learning across CNNs for emotion recognition through visual cues, while in \cite{kollias2017recognition}, the authors combine CNNs and recurrent neural networks for visual-based arousal and valence modeling. 

All the methods discussed above exploit data that either has been collected in well-defined and controlled laboratory conditions or directly depict the facial expressions of a human subject whose emotional state needs to be predicted. Collecting data in a laboratory is an invasive task that requires specialized software and hardware, limiting this way the application of affect models in real-life scenarios. At the same time, directly depicting human subjects raises privacy and personal data protection issues. This study aims to move affective modeling outside of a laboratory's closed boundaries by building models able to predict affect using information that is available in the wild. Moreover, it adopts a user-agnostic perspective using solely the visual information of human-computer interaction for predicting humans' emotional states.

\subsection{Affect Modeling In The Wild}

Affect modeling in the wild focuses on developing models able to analyse the emotional state of humans in real-life scenarios that entail uncontrolled conditions. Towards this direction large databases \cite{zafeiriou2017aff,kollias2019expression,mavadati2013disfa,zhang2014bp4d,zheng2018multimodal} that simulate human emotions in the wild are necessary \cite{kollias2016line}. Having large affect corpora available enables the development of powerful deep learning models that achieve state-of-the-art results. Indicatively, AffWildNet \cite{kollias2017recognition,kollias2019deep} effectively combines CNNs and recurrent neural networks to accurately capture the face dynamics and achieve the best performance in \cite{zafeiriou2017aff}. Toisoul \textit{et al.} \cite{toisoul2021estimation} present the EmoFAN deep learning model that builds on top of the face alignment network \cite{bulat2017far} to predict jointly discrete emotional states and continuous affect dimensions. EmoFAN achieves the best performance on the AfewVA dataset \cite{kossaifi2017afew}. Aspandi \textit{et al.} \cite{aspandi2020adversarial} use adversarial-based neural networks to learn latent representations from audiovisual signals for estimating affect in the wild. The authors of \cite{parthasarathy2021detecting} use multimodal transformers to capture and exploit temporal dynamics of audiovisual information towards detecting affect states. They demonstrate that multimodal deep learning affect models can significantly improve affect detection in the wild. Finally, Kolias and Zafeiriou \cite{kollias2021affect} propose a unified framework for affect modeling in the wild that considers facial expressions and categorical affect, facial action units, and dimensional affect representations. 

Although the studies listed above model affect using information captured in natural conditions, they all require direct measurements from humans in the wild. Measurements such as facial expressions employ sensitive personal data handling that can limit the derived affect models' application in real-life scenarios. As mentioned above, this study follows a subject-agnostic approach in affect modeling by using visual information of human-computer interaction as sole input for affect prediction. This crucial difference eliminates any privacy issues, enabling our models' unrestricted application in the wild and in real-world scenarios.  

\section{Datasets}

\begin{figure}[!tb]
	\begin{minipage}{0.49\linewidth}
		\centering
		\centerline{\includegraphics[width=0.98\linewidth]{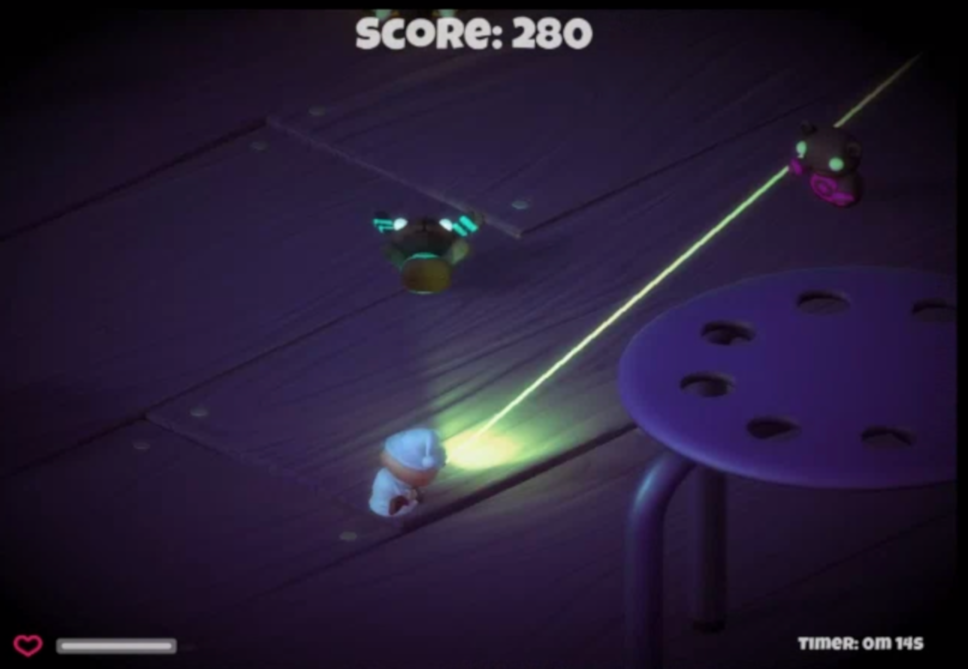}}
	\end{minipage}
	\begin{minipage}{0.49\linewidth}
		\centering
		\centerline{\includegraphics[width=0.98\linewidth]{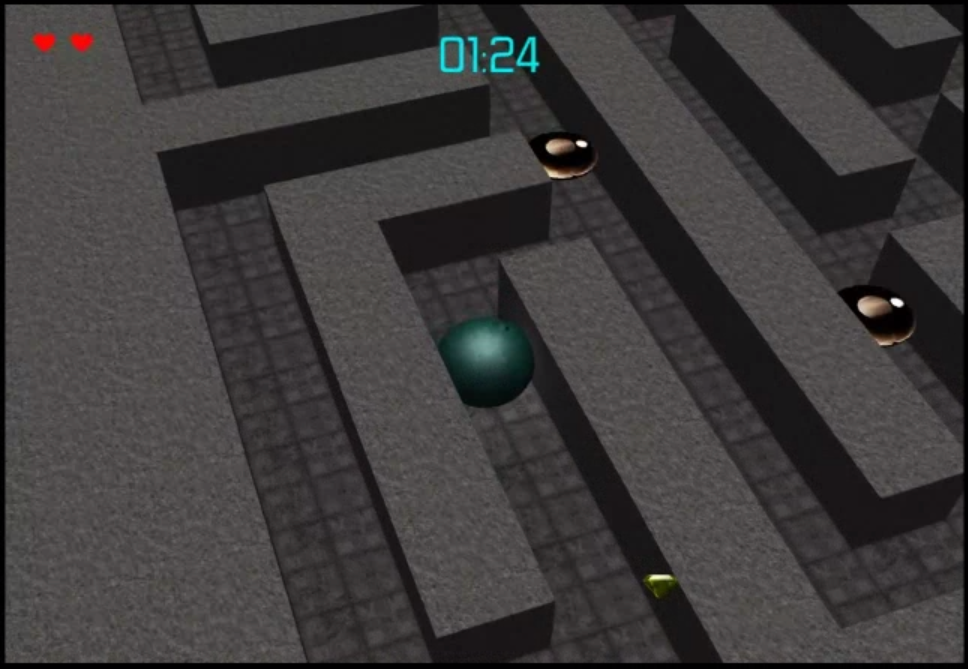}}
	\end{minipage}
	\caption{Screenshots from Survival Shooter (left) and Space-Maze (right) games.}
	\label{fig:game_screenshots}
\end{figure}

To test the impact of privileged information on modeling players' affect, we selected two dissimilar games: Survival Shooter and Space Maze depicted in Fig. \ref{fig:game_screenshots}. These games belong to different genres and feature different mechanics, pace and visual design. Survival Shooter is a fast-paced game that requires accurate aiming and constant movement. Space Maze, on the other hand, is a slow-paced physics game that requires accurate movement timing. We use these two games as our initial test-beds for investigating the degree to which LUPI is beneficial for affect modelling. 

\subsection{Survival Shooter}

Survival Shooter (SS) \cite{camilleri2017towards} is a game adapted from a tutorial package of Unity3D. The player has to shoot down as many hostile toys as possible and avoid collisions with them. The hostile toys spawn at predetermined areas of the level and move towards the player's avatar, which is equipped with a laser gun for killing them. The maximum duration of gameplay is 60 seconds. 

\subsection{Space Maze}

Space Maze (SM) is a 3D maze-based puzzle game \cite{yannakakis2010towards, knight2013space}. The player controls a cyan ball in a maze that contains dark ball-shaped enemies and three diamond-shaped tokens. A player has to collect all three diamond tokens and move the cyan ball to a predefined goal point within 90 seconds, without running out of health due to collisions with enemies.

\subsection{Experimental Protocol}

This study's SS and SM data was collected from 25 players (10 females) aged from 19 to 54 (median age 24); 70\% of the participants considered themselves good or expert players, while 30\% considered themselves novice or non-gamers. The participants were recruited via snowball sampling and were primarily university students with no prior experience in affect annotation. Before annotation, all participants were presented with an introductory screen that describes arousal as ``the intensity of gameplay no matter whether you like the game or not. High arousal can be a feeling of readiness, tension, excitement or exhilaration. Low arousal can be a feeling of fatigue, boredom, calmness or relaxation''. 

\begin{figure}[!tb]
    \centering
    \includegraphics[width=1\linewidth]{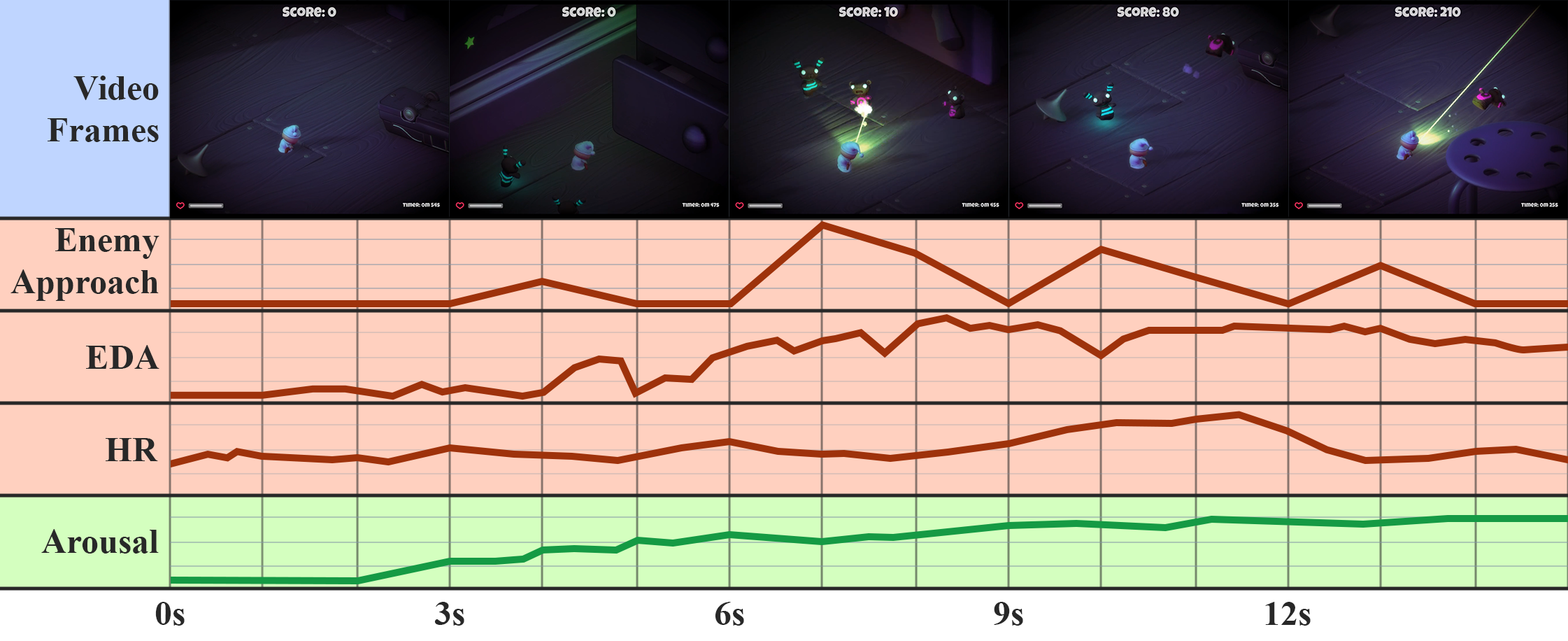}
    \caption{An example of the user modalities considered and the corresponding arousal trace over time. In blue: extracted frames. In red: physiology (EDA, HR) and an indicative telemetry feature ($E_a$). In green: annotation trace. Red indicates privileged information in this paper.}
    \label{fig:features}
\end{figure}

Each participant played a game and then annotated her recorded gameplay footage in terms of arousal using the RankTrace tool \cite{lopes2017ranktrace, melhart2019pagan}, which allows continuous and unbounded annotations. This play-annotation cycle occurred twice for each of the games, resulting in 50 annotated gameplay videos for each game. While the participants played the games, the Empatica E4 wristband was fitted on their left wrist for logging Heart Rate (HR) and Electrodermal Activity signals (EDA). HR and EDA were captured at 1Hz and 4Hz, respectively, while the gameplay footage was recorded at 30 frames per second (30Hz), and RankTrace provided 4 annotations per second (4Hz). Finally, along with the gameplay footage, we logged gameplay telemetry data describing the main events that happened during the gameplay. Figure~\ref{fig:features} shows the different input modalities of the dataset and the corresponding output (\emph{arousal}).

\subsection{Data Preprocessing and Feature Extraction}
\label{ssec:data_preprocessing}

This study aims to produce models of affect that predict arousal based on three different information streams: frames of gameplay footage, gameplay telemetry data, and physiological measurements. We split each gameplay session using overlapping time windows. The sliding step, as well as the length of the windows are hyperparameters. In this study, we conduct experiments for 0.5 seconds sliding step and 1, 3, and 5 seconds window length. Using a fixed sliding step and overlapping time windows, the dataset size (number of time windows) is not affected by windows' length. By varying the windows' length, the amount of temporal information encoded in each window changes affecting both visual information and telemetry/physiology features.   

Since RankTrace provides continuous and unbounded annotations, we first normalize the arousal annotations to a $[0,1]$ value range in a game session-wise manner. Each window is described by the visual information, the gameplay telemetry and physiology data, and the normalized arousal annotations captured during its duration. The visual information is represented by a sequence of scaled RGB frames ($160 \times 90$ pixels) concatenated along the channels' dimension. The physiology features correspond to the average and average gradient of HR and EDA measurements, while the arousal label for a window corresponds to the average of the normalized arousal annotations.
We extract 23 features from the gameplay telemetry data. These features can be divided into four main categories: features describing the overall \emph{gameplay} context, 
features relating to the \emph{player}, 
features relating to game \emph{objects}, 
and features related to \emph{enemies}. 
Most features in the dataset represent the frequency of corresponding events within a time-window. 
Table~\ref{tab:features} summarizes the telemetry features extracted with their corresponding descriptions.
%


\begin{table}[!tb]
    \centering
    \caption{List of telemetry features used in the dataset. Type indicates whether the feature is associated with  Gameplay (G), Player (P), Objects (O) or Enemies (E)}
    \begin{tabular}{c|c|l}
    \textbf{Type} & \textbf{Feature} & \textbf{Description} \\
    \hline
    \hline
    \multirow{3}{*}{G} & $H_h$ & Heuristic about helpful events \\
    & $H_d$ & Heuristic about detrimental events \\
    & $T$ & time since level started \\
    \hline
    \multirow{12}{*}{P} & $P_M$ & player movement \\
    & $D$ & Distance from last position \\
    & $T_M$ & Percentage of time spent moving \\
    & $P_{AG}$ & Player moves away from the goal \\
    & $P_{TG}$ & Player moves towards the goal \\
    & $P_A$ & Player attacks \\
    & $P_H$ & Player hits an enemy \\
    & $P_K$ & Player kills an enemy \\
    & $P_h$ & Player recovers health \\
    & $P_d$ & Player discovers new area \\
    & $P_C$ & Player collects a pickup \\
    & $P_W$ & Player wins \\
    \hline
    \multirow{2}{*}{O} & $P_a$ & Pickup appears on screen \\
    & $P_d$ & Pickup disappears from screen \\
    \hline
    \multirow{6}{*}{E} & $E_a$ & Enemy appears on screen \\
    & $E_d$ & Enemy disappears from screen \\
    & $E_e$ & Enemy close to player (heuristic) \\
    & $E_c$ & Enemy starts chase \\
    & $E_h$ & Enemy hits player \\
    & $E_k$ & Enemy kills player \\
    \end{tabular}
    \label{tab:features}
\end{table}

\section{Affect Modeling Using Privileged Information}

In this section we detail the \emph{Learning Using Privileged Information} paradigm \cite{vapnik2009new} for building models of affect capable of generalizing in the wild, as well as the architecture of the employed machine learning models.

\subsection{Learning Using Privileged Information}

Vapnik formally introduced the Learning Using Privileged Information (LUPI) paradigm in \cite{vapnik2009new, vapnik2015learning}. LUPI targets problems characterized by an asymmetric distribution of information between training and test time; that is, additional information is given about the training data, which is not available at test time. This setting is prevalent in affective computing. In laboratory conditions, different modalities of information can be captured to model humans' affect. In the wild, however, it is very difficult or even impossible to capture the same modalities due to the sensing devices' cost and the invasiveness of the capturing procedures.

LUPI provides the means to \textit{transfer knowledge} from all the available modalities to a machine learning model that makes predictions using only a subset of these modalities \cite{sharmanska2013learning,lopez2016unifying}. In other words, LUPI allows a machine learning model of affect to be trained exploiting information that comes from all the modalities captured in a laboratory setting. During test time, however, the same model makes predictions using only those modalities that are available in the wild. The information that is not available during test time is called \emph{privileged information}. In our dataset, we consider as privileged the information that comes from gameplay telemetry and physiology measurements since this kind of information requires special laboratory hardware and software to be captured. On the contrary, we consider that visual information (captured using only a screen recorder) is available both at training and test times. In the following, we describe transferring knowledge from privileged information to a machine learning model. At this point, we should clarify that transferring knowledge using LUPI is different from the transfer of learning techniques used in deep learning \cite{ng2015deep}. Transfer of learning targets small-sample setting problems by finetuning a model trained for a specific task such that it performs well in a similar task. On the contrary, using LUPI focuses on problems with asymmetric distribution of training/testing information and trains the student model from scratch. 

Before transferring knowledge that comes from privileged information, we first have to represent it appropriately. Following \cite{hinton2015distilling,lopez2016unifying}, we represent that knowledge within the probabilistic predictions of a trained machine learning model and make predictions using only privileged information. This model is called \emph{teacher} and, under a classification setting, it is trained following the typical supervised learning paradigm, i.e. by minimizing the cross-entropy loss
\begin{equation}
L_{CE}(p_n) = - \frac{1}{N} \sum_{n=1}^N y_n^T \log(p_{n}),
\end{equation}
where $N$ is the number of samples in the training set, $y_n$ is the ground truth label (one-hot encoded) of the $n$-th sample, and $p_{n}$ is the probabilistic output of the teacher model.

Having a teacher model trained, we can transfer the knowledge from privileged information to another model, called \emph{student}. The student model makes predictions based only on the information that is available in the wild. Based on \cite{hinton2015distilling} and \cite{lopez2016unifying}, the transfer of knowledge can be achieved by feeding the student only with those modalities of information available in the wild and force it during training to balance between hard ground truth labels and soft teacher's probabilistic predictions. This balancing can be formally defined in the following loss function
\begin{equation}
\label{eq:student_loss}
L_{student} = (1 -\alpha) L_{CE}(q_n) + \alpha L_{KL},
\end{equation}
where 
\begin{equation}
L_{KL} = \frac{1}{N} \sum_{n=1}^N q_n \log(\frac{q_n}{p_n})
\end{equation}
is the Kulback-Leibler divergence loss, $\alpha \in [0,1]$, and $q_n$ (a vector with positive elements whose sum equals 1) is the probabilistic prediction of the student for the $n$-th sample. In the presented case study, we train the teacher using gameplay telemetry and physiology features, while the student is trained based on gameplay footage frames and the teacher's predictions. We should emphasize that after training, the student model makes predictions using information \emph{solely} from gameplay footage frames. 

\begin{figure*}[!tb]
	\begin{minipage}{1.0\linewidth}
		\centering
		\centerline{\includegraphics[width=0.7\linewidth]{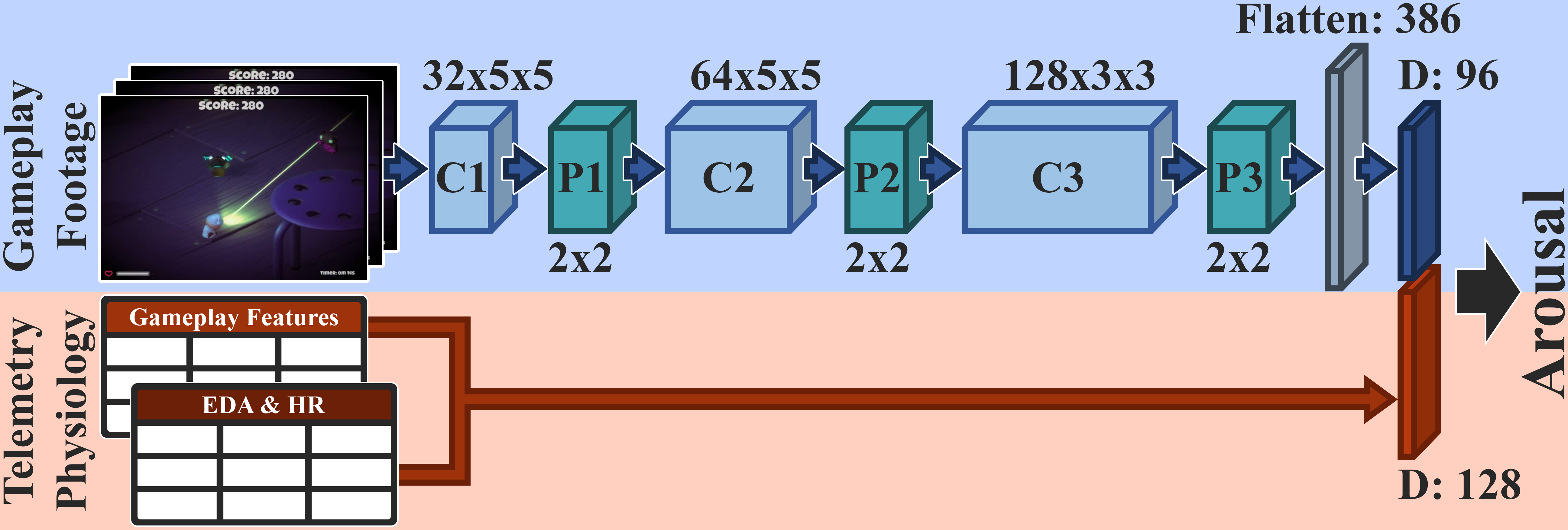}}
	\end{minipage}
	\caption{Architecture of the employed machine learning models of affect. The convolutional, max-pooling and dense layers are denoted by ``C'', ``P'' and ``D'', respectively.}
	\label{fig:architecture}
\end{figure*}

\subsection{Machine Learning Models of Affect}

We build two machine learning models of affect: one student and one teacher. The teacher model is a fully connected feedforward neural network with one hidden layer that contains 128 neurons. The teacher network uses privileged information---gameplay telemetry and physiology features---during training and test time. We fuse privileged information following an early fusion approach by concatenating the telemetry and physiology vectors before using them as input to the model. The architecture of the teacher network corresponds to the red stream in Fig. \ref{fig:architecture}. The student model is a 2D convolutional neural network with three convolutional layers that receives as input video frames concatenated along the channels dimension. The first two convolutional layers consist of 32 and 64 learnable kernels of dimension $5 \times 5$ and stride equal to 2. The third convolutional layer consists of 128 learnable kernels of dimension $3 \times 3$, and stride of 1. A $2 \times 2$ max-pooling layer follows each of the convolutional layers. The last convolutional layer's output is fed to a dense layer with 96 hidden neurons and then is propagated to the output layer. At the penultimate layer of the student network we also use dropout with a probability parameter of 0.1. For both models we use Adam optimizer with learning rate 0.001. The architecture of the student model corresponds to the blue stream in Fig. \ref{fig:architecture}. Finally, we use ReLU as the activation function for all layers,  for both teacher and student models.

To evaluate the impact of privileged information on affect models' performance, we build two more models that serve as baselines. The first is a convolutional neural network that is being trained and makes predictions by exploiting only the information of gameplay pixels available on the footage frames; we name this model \emph{pixel-based} following the reported benefits of modeling affect solely from pixels \cite{makantasis2019pixels, makantasis2021pixels}. The architecture of this model is the same as the architecture of the student model and visually presented by the blue stream in Fig. \ref{fig:architecture}. The second model, namely \emph{fusion}, fuses and exploits all modalities (privileged information and gameplay footage frames) both during training time and for making predictions. We fuse information from different modalities following a late fusion approach. The stream that processes the visual information has the same architecture as the student model, while the stream that processes the telemetry/physiology features has the same architecture as the teacher model. The learned representations of the two streams are concatenated to form a fused representation, which is directly propagated to the output layer for training and making predictions. The architecture of the fusion network corresponds to both streams in Fig. \ref{fig:architecture}. 

\section{Results}

This section presents the training data preparation, the framework for evaluating the impact of privileged information on affect modeling, and the experimental results. 

\subsection{Training Data and Evaluation Framework}

In this study, we evaluate the impact of privileged information on affect modeling in terms of arousal prediction accuracy. 
The RankTrace annotation tool provides continuous and unbounded values of arousal, and thus it may seem natural to view the arousal prediction problem as a regression task. We wish, however, to investigate the performance of affect models under a user-agnostic approach without making any assumptions regarding the value of the annotations, which, in turn, result in biased and user-specific models \cite{yannakakis2018ordinal}. For this reason, we view arousal prediction as a binary classification task \cite{makantasis2021pixels}---labels of low and high arousal---and define arousal prediction accuracy in terms of binary classification accuracy. 

We use the approach presented in \cite{makantasis2021pixels} to transform the normalized arousal value of a gameplay time window into a low or high arousal label. 
In particular, we compute the mean value of the normalized annotations within each game session. If the normalized annotation of the examined gameplay time window is larger than the mean value of its session plus a threshold $\epsilon$, we assign to that window the label of high arousal. Similarly, if its normalized annotation value is lower than the session mean value minus the threshold $\epsilon$, we assign to it the low arousal label. The $\epsilon$ parameter determines a region around the mean within which annotation values are labelled as uncertain and ignored during classification to avoid unstable classifiers due to trivial differences in their inputs. Based on the successful findings of \cite{makantasis2021pixels}, we set $\epsilon=0.1$.

\begin{table}[!tb]
\centering
\caption{Dataset sizes for the SS and SM games across different gameplay windows' length.}
\begin{tabular}{p{2.5cm}|>{\centering\arraybackslash}p{1.5cm}|>{\centering\arraybackslash}p{1.5cm}|>{\centering\arraybackslash}p{1.5cm}}
\hline \hline
     & 1 second & 3 seconds & 5 seconds  \\ \hline
Survival Shooter (SS) & 3221 & 3039  & 2845  \\ 
Space Maze (SM)   & 4379 & 4174  & 3967    \\ \hline \hline
\end{tabular}
\label{table:dataset_sizes}
\end{table}

Following the approach described in Section \ref{ssec:data_preprocessing} and the binary labels transformation mentioned above, we construct the datasets to train the affect models. As mentioned before, to split the gameplay sessions, we use overlapping windows of length 1, 3, and 5 seconds with sliding step 0.5 seconds. Table \ref{table:dataset_sizes} presents the cardinality of the datasets for the two games and different windows' length. To evaluate models' performance, we follow a 5 fold cross-validation scheme. When splitting the dataset, we do not include sessions from the same player in both training and testing data. We also use $10\%$ of the training data as a validation set to activate early stopping criteria and avoid models overfitting. Specifically, the training stops after 15 epochs without loss improvement on the validation set. At this point, we should emphasize that all the employed models of affect are evaluated using precisely the same data, i.e. the training, validation and test sets are the same for all models.

\subsection{Teacher's Impact on Student's Performance}

We start by investigating the impact of the teacher on the performance of the student model as determined by the parameter $\alpha$ in Eq.~\eqref{eq:student_loss}. Initially, we train the teacher using privileged information for gameplay footage windows of 1, 3, and 5 seconds. After training the teacher models, we use their probabilistic predictions to train the student models fed only with gameplay footage frames. We train the student models using three different values for parameter $\alpha$: $\alpha=0.2$, $0.5$, and $0.8$. By increasing the value of $\alpha$, we force the student model to weigh more the predictions of the teacher and pay less attention to ground truth labels.  

Table \ref{table:alpha} presents the results of this investigation. For both datasets, there is a value for parameter $\alpha$, which results in student models that achieve higher accuracy than the teacher, irrespectively of the gameplay footage window length. We observe, however, that varying the value of parameter $\alpha$ yields fluctuations in the performance of the student, especially in longer time windows. For $\alpha=0.5$ and $0.8$, the student achieves the best accuracy, i.e. when the teacher's predictions are weighted more than or equal to the ground truth labels. That indicates that privileged information encoded in teacher's predictions can effectively improve the students' training.


\begin{table}[tb]
\centering
\caption{The effect of $\alpha$ parameter on students' average accuracy (\%) across the two datasets and three time windows.}
\begin{tabular}{p{2.2cm}|>{\centering\arraybackslash}p{1.6cm}|>{\centering\arraybackslash}p{1.6cm}|>{\centering\arraybackslash}p{1.6cm}}
\multicolumn{4}{c}{\textbf{Survival Shooter}}  \\ \hline \hline
& 1 second & 3 seconds & 5 seconds \\ \hline
Majority Class & 52.5 & 52.9  & 53.3  \\ 
Teacher    & 70.0 & 68.8  & 73.4  \\ \hline
Student ($\alpha=0.2$)    & 67.0 & 65.2  & 66.6  \\ 
Student ($\alpha=0.5$)   & \textbf{71.2} & \textbf{74.3}  & 71.1  \\ 
Student ($\alpha=0.8$)   & 68.5 & 70.2  & \textbf{74.3}  \\ \hline \hline 

\multicolumn{4}{c}{} \\
\multicolumn{4}{c}{\textbf{Space Maze}}  \\ \hline \hline
& 1 second & 3 seconds & 5 seconds \\ \hline
Majority Class  & 52.7 & 52.9  & 52.3  \\ 
Teacher  & 71.9 & 73.3  & 70.8 \\ \hline 
Student ($\alpha=0.2$)  & 75.6 & 66.0  & 56.4 \\ 
Student ($\alpha=0.5$)  & 74.8 & \textbf{75.1}  & 62.9 \\ 
Student ($\alpha=0.8$)  & \textbf{76.4} & 70.1  & \textbf{73.5}  \\ \hline \hline 
\end{tabular}
\label{table:alpha}
\end{table}


Based on these results, we can conclude that $\alpha$ is an important parameter, which, when appropriately set, yields student models that outperform the teacher. Most importantly, student models achieve that level of improvement without access to privileged information; they make predictions exploiting only that information available in the wild. In other words, the student models can take advantage of privileged information captured in laboratory conditions to generalize in real-life scenarios where privileged information is not available.

\subsection{The Importance of Privileged Information}

\begin{figure}[!tb]
	\begin{minipage}{1.0\linewidth}
		\centering
		\centerline{\includegraphics[width=1.0\linewidth]{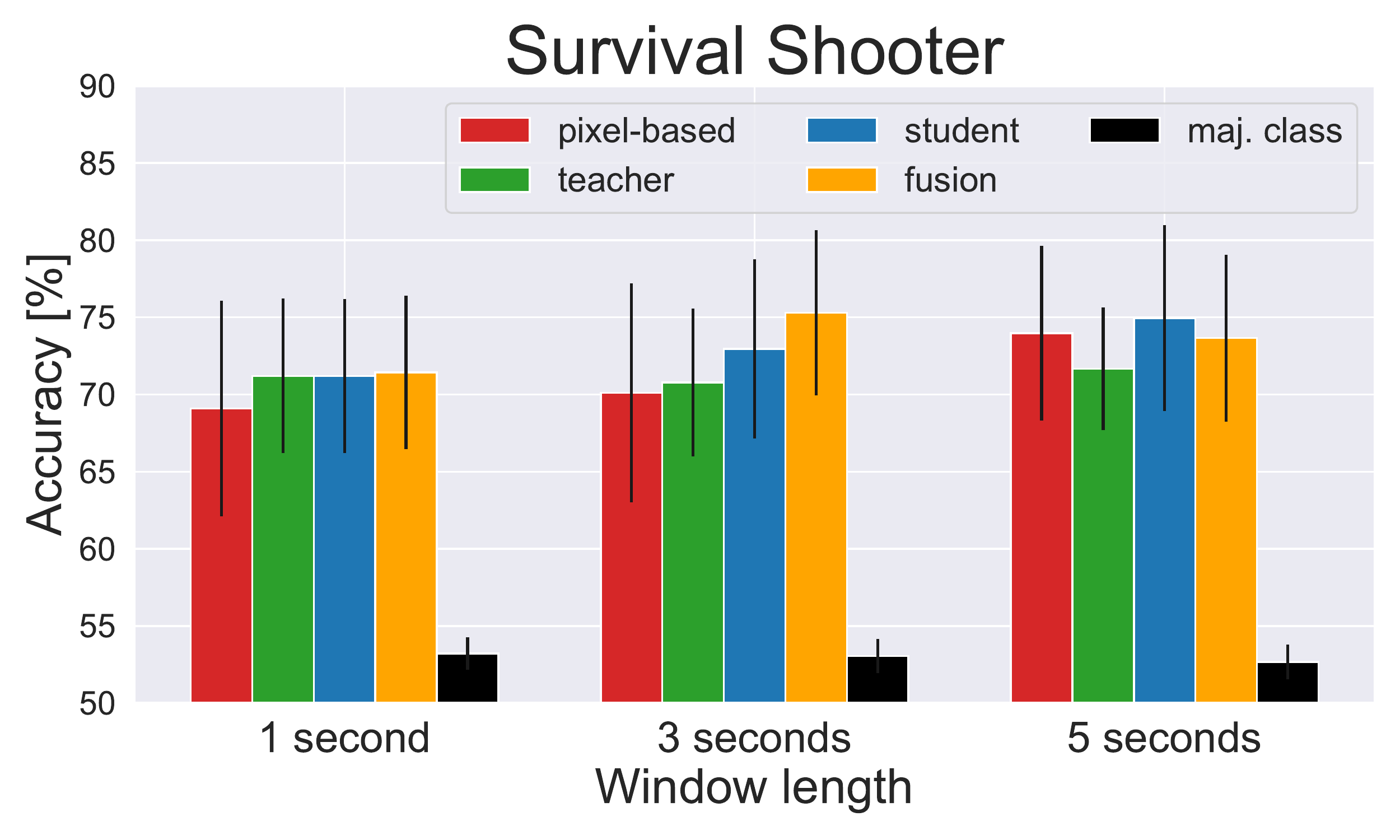}}
	\end{minipage}
	\begin{minipage}{1.0\linewidth}
		\centering
		\centerline{\includegraphics[width=1.0\linewidth]{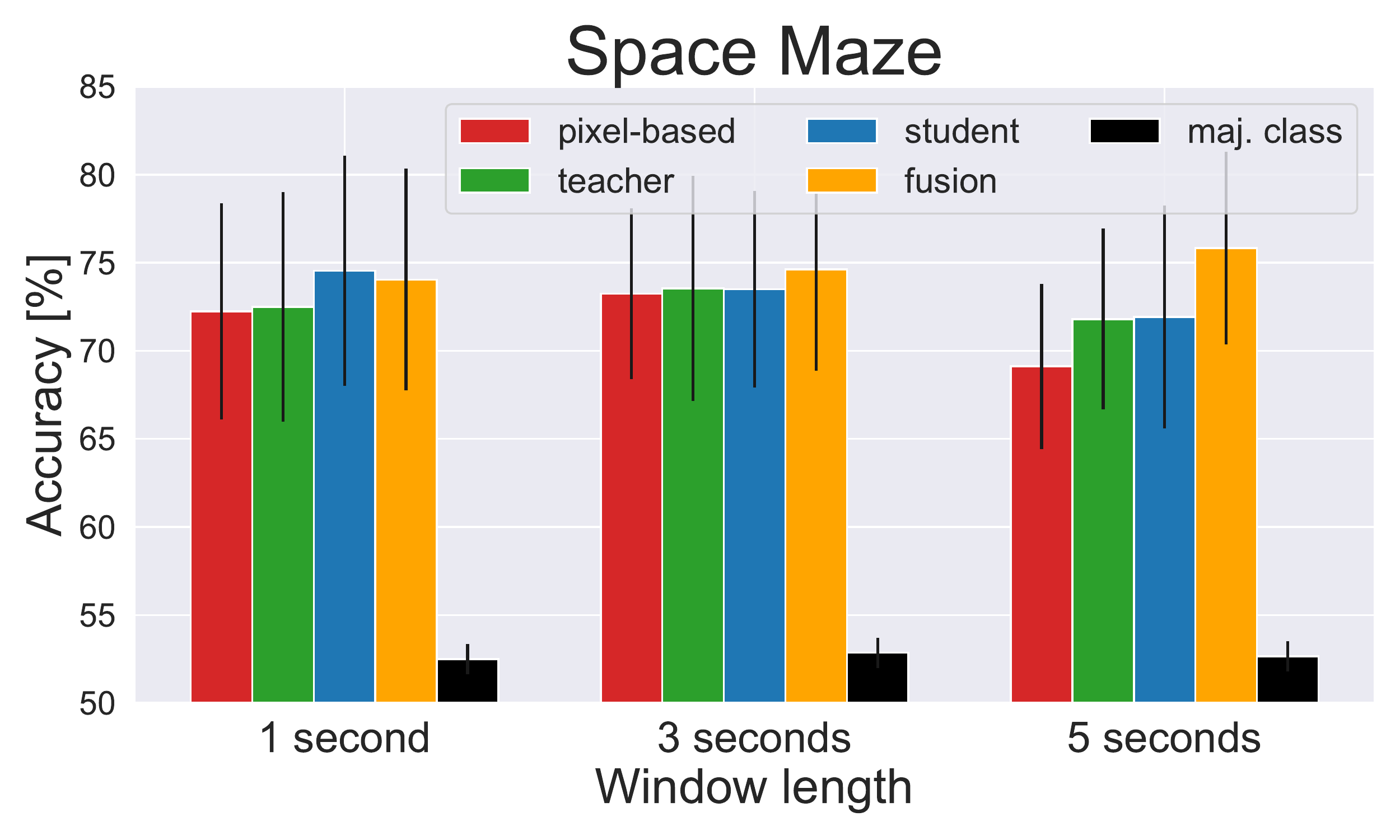}}
	\end{minipage}
	\caption{Comparison of the employed models on the two datasets and different time windows, in terms of average accuracy and 95\% confidence intervals from running 3 times the 5-fold cross-validation scheme.} 
	\label{fig:comparison}
\end{figure}

In a second set of experiments, we investigate the impact of privileged information on building accurate models of affect. As mentioned before, the student models make predictions using solely information that is available in the wild; in our case, the frames of gameplay footage that we process as RGB pixels. We first compare the student models' accuracy against the accuracy of a pixel-based model that uses as input the same information as the student models. Second, and more importantly, we compare student models built to operate in the wild against a \emph{fusion model} that uses all information modalities captured in laboratory environments for training and testing. 
For the following experiments, we use $\alpha$ parameter values that yield the most accurate student models based on the sensitivity analysis of the previous section. Also, we run 3 times the 5-fold cross-validation scheme, with different training/validation/testing data splits and models initialization, to collect statistics regarding the performance of the models. As a baseline performance, we also report the accuracy of a dummy classifier, denoted as \textit{majority class}, which always outputs the most frequent class in the training set. The results for these comparisons are presented in Fig. \ref{fig:comparison}.

For the SS dataset, the fusion model achieves the best accuracy for two out of the three gameplay footage window lengths, specifically for 1 and 3 seconds. For 5 second windows, the student model performs better than all other models on average, followed by the pixel-based model. The accuracy, however, of the pixel-based model seems highly dependent on the window's length since we observe large fluctuations between 1, 3, and 5 second windows. For all window lengths, the student is the only model that consistently achieves accuracy very close to (or even better than) that of the fusion model, despite using only RGB gameplay footage pixels for making predictions. At the same time, and in contrast to the pixel-based model, the student's accuracy is robust to window length changes. While the pixel-based and the student models use the same kind of information for making predictions, the latter exploits privileged information and achieves higher accuracy on average across all 3 different window lengths. 

For the SM dataset and all window lengths, the fusion and the student are the two best performing models. The student model achieves higher accuracy than the fusion model for 1 second window length, despite utilising only a subset of the modalities of the fusion model.
The performance of the pixel-based model presents smaller fluctuations compared to the SS dataset. However, it achieves lower accuracy than the student model for all settings, even though both models use the same information for making predictions. 




To summarize, the student is the only model that consistently achieves average accuracy values close to or even higher (for 2 out of 6 scenarios) than those of the fusion model. We observe that behaviour across all scenarios tested, indicating that the LUPI paradigm can provide the means for building accurate models of affect that operate in the wild. Moreover, the student and the pixel-based models use the same kind of information for making predictions. The student, however, appears more robust across different scenarios and achieves higher average accuracy for most settings.

\section{Discussion}

Testing affect models in the wild comes with costs associated primarily to affect sensing. One would assume that if an affect model has access to fewer modalities during testing in the wild (e.g. due to hardware/software failure or even due to the unavailability of sensors) the result will be detrimental for its accuracy. Surprisingly, however, the results obtained from this initial study suggest otherwise. Learning using privileged information via the LUPI paradigm \cite{vapnik2015learning} seems to enable affect models to operate in the wild---having access only to a limited set of modalities; in this study merely pixels---without an actual cost in terms of performance. Our findings suggest that LUPI models can perform equally well to fusion models that consider all modalities during both training and testing. Most importantly via LUPI affect models we gain on accessibility, cost, intrusiveness, and privacy, bringing affective computing a decisive step closer to real-world settings.

A key characteristic of the domain of games generally---and the two games tested specifically---is that the modality of gameplay footage is a very powerful predictor of arousal, as indicated by earlier studies \cite{makantasis2019pixels,makantasis2021pixels}. As a result the student model benefits to a certain yet limited degree by additional modalities. Results obtained in this paper show that pixel-based arousal models in some experimental settings appear to perform equally well, on average, to the student model or even to the fusion models. This domain characteristic does not undervalue the contribution of LUPI, however, as student models yield on average higher or similar accuracies to pixel-based models. Most importantly for the proposes of this study, the student models perform equally well to the fusion models. 

While the LUPI method appears to be robust across the modalities and test beds examined in this paper, our hypothesis that privileged information is beneficial for multimodal affect-based interaction needs to be tested further. In particular, we plan to employ variations of the method across different affective corpora---within games \cite{melhart2021again} and beyond \cite{koelstra2011deap}---that contain dissimilar modalities of user input and various emotional labels. Even though standard convolutional neural networks appear to be performing well in this and earlier studies \cite{makantasis2019pixels,makantasis2021pixels}, our plan is to test a number of different deep learning models for potentially improving the performance of LUPI models. Another possible extension of this work is to test and compare ordinal learning methods \cite{yannakakis2018ordinal,yannakakis2017ordinal} for deriving models of affect through privileged information. 

\section{Conclusions}

In this paper we introduce the notion of \emph{privileged information} for building models of affect. Our hypothesis is that learning using privileged information can aid affect models to leave the lab, be tested and perform well in real-world settings. To test our hypothesis we used an affect corpus from two different game environments that contains 4 modalities of user input: game telemetry data, heart rate, electrodermal activity and the pixels of the gameplay footage. We consider the first 3 modalities as privileged information (i.e. only available during data collection) and assume that the fourth modality (i.e. pixels) is available both during training and testing. The core results of this initial study suggest that arousal models trained with the privileged information of the three user modalities can ignore them during testing and perform equally well to fusion arousal models that consider all modalities. The findings of the paper bring affective computing one step closer to realising affect interaction in the wild. Privileged information affect models do not require access to costly, intrusive, or impractical modalities when tested in the wild, and they can still operate equally well to models that consider all this additional information.

The proposed method has direct applications to any affect modelling task that considers multimodal data and needs to run in the wild. Potential applications include (but not limited to) driver-assistive systems, affective robots, affect-aware recommender systems, and health applications at home such as stress monitoring and seizure detection.

\bibliographystyle{IEEEbib}

\end{document}